\documentclass[prb,twocolumn,showkeys,showpacs,amsmath,amsfonts,floatfix,superscriptaddress,nofootinbib]{revtex4-1}

\usepackage{amssymb,amsmath,amstext}                
\usepackage{graphicx}                                               
\usepackage{epstopdf}                                               
\usepackage{color}       
\usepackage{bm}
\usepackage{appendix}                                              
\usepackage[latin2]{inputenc}
\usepackage{ulem}
\usepackage{latexsym}
\usepackage{hyperref}
\def\be{\begin{equation}}
\def\ee{\end{equation}}
\def\bea{\begin{eqnarray}}
\def\eea{\end{eqnarray}}
\def\bi{\begin{itemize}}
\def\ei{\end{itemize}}

\begin{document}

\title{ Time Evolution of an Infinite Projected Entangled Pair State:\\
                       an Efficient Algorithm     }

\author{ Piotr Czarnik }
\affiliation{Institute of Nuclear Physics, Polish Academy of Sciences, 
             Radzikowskiego 152, PL-31342 Krak\'ow, Poland}
             
\author{ Jacek Dziarmaga }
\affiliation{ Marian Smoluchowski Institute of Physics, Jagiellonian University,
              ul. Prof. S. {\L}ojasiewicza 11, PL-30348 Krak\'ow, Poland }             
             
\author{ Philippe Corboz }
\affiliation{Institute for Theoretical Physics and Delta Institute for Theoretical Physics, 
             University of Amsterdam, Science Park 904, 1098 XH Amsterdam, The Netherlands}                    
             
\date{\today}

\begin{abstract}
An infinite projected entangled pair state (iPEPS) is a tensor network ansatz to represent a quantum state on an infinite 2D lattice whose accuracy is controlled by the bond dimension $D$. Its real, Lindbladian or imaginary time evolution can be split 
into small time steps. Every time step generates a new iPEPS with an enlarged bond dimension $D' > D$, which is approximated by an iPEPS  with the original $D$. 
In Phys. Rev. B 98, 045110 (2018)
an algorithm was introduced to optimize the approximate iPEPS by 
maximizing directly its fidelity to the one with the enlarged bond dimension $D'$. 
In this work we implement a more efficient optimization employing a local estimator of the fidelity. 
For imaginary time evolution of a thermal state's purification, 
we also consider using unitary disentangling gates acting on  ancillas  to reduce the required $D$.  
We test the algorithm simulating Lindbladian evolution and unitary evolution 
after a sudden quench of transverse field $h_x$ in the 2D quantum Ising model.
Furthermore, 
we simulate thermal states of this model and estimate the critical temperature with good accuracy:
$0.1\%$ for $h_x=2.5$ and $0.5\%$ for the more challenging case of $h_x=2.9$ close to the quantum critical point at $h_x=3.04438(2)$.
\end{abstract}

\maketitle

\section{Introduction}
\label{sec:introduction}

Weakly entangled quantum states of strongly correlated systems can be efficiently represented by tensor networks\cite{Verstraete_review_08,Orus_review_14}. The most popular are a 1D matrix product state (MPS) \cite{Fannes_MPS_92} 
and its 2D generalization: pair-entangled projected state (PEPS) \cite{Verstraete_PEPS_04}. 
An MPS provides a compact representation of ground states of gapped local Hamiltonians \cite{Verstraete_review_08,Hastings_GSarealaw_07,Schuch_MPSapprox_08} and 
purifications of thermal states of local Hamiltonians \cite{Barthel_1DTMPSapprox_17}. 
It is the ansatz optimized by the density matrix renormalization group (DMRG) \cite{White_DMRG_92, White_DMRG_93,Schollwock_review_05,Schollwock_review_11}. 
In 2D PEPS are expected to be efficient representation of ground states of gapped local Hamiltonians \cite{Verstraete_review_08,Orus_review_14} and thermal states of 2D local Hamiltonians \cite{Wolf_Tarealaw_08,Molnar_TPEPSapprox_15}, 
though in 2D the representability of area-law states by tensor networks was demonstrated to have limitations\cite{Eisert_TNapprox_16}.
2D tensor networks can represent fermionic systems \cite{Corboz_fMERA_10,Eisert_fMERA_09,Corboz_fMERA_09,Barthel_fTN_09,Gu_fTN_10}, 
as was shown for both finite \cite{Cirac_fPEPS_10} and infinite PEPS \cite{Corboz_fiPEPS_10,Corboz_stripes_11}. 

Originally, PEPS was proposed as an ansatz for ground states of finite 2D systems \cite{Verstraete_PEPS_04,Murg_finitePEPS_07} generalizing earlier attempts to construct trial wave-functions for specific 2D models \cite{Nishino_2DvarTN_04}. 
Efficient numerical methods for infinite PEPS (iPEPS) \cite{Cirac_iPEPS_08,Xiang_SU_08,Gu_TERG_08,Orus_CTM_09} 
made it a promising new tool for strongly correlated systems. Its achievements include solution of a long standing magnetization plateaus problem in the highly frustrated compound $\textrm{SrCu}_2(\textrm{BO}_3)_2$ \cite{Matsuda_SS_13,Corboz_SS_14} and demonstrating that the ground state of the doped 2D Hubbard model has stripe order   \cite{Simons_Hubb_17}. 
Another example is the kagome Heisenberg antiferromagnet for which new evidence supporting gapless spin liquid was obtained  \cite{Xinag_kagome_17}. 
This progress was made possible by new developments in 
iPEPS optimization \cite{fu,Corboz_varopt_16,Vanderstraeten_varopt_16}, 
contraction \cite{Fishman_FPCTM_17,Xie_PEPScontr_17,Czarnik_fVTNR_16}, 
energy extrapolations \cite{Corboz_Eextrap_16}, 
and universality class estimation \cite{Corboz_FCLS_18,Rader_FCLS_18,Rams_xiD_18}. 

These achievements encourage to use iPEPS for a broad class of 2D states like 
thermal states \cite{Czarnik_evproj_12,Czarnik_fevproj_14,Czarnik_SCevproj_15, Czarnik_compass_16,Czarnik_VTNR_15,Czarnik_fVTNR_16,Czarnik_eg_17,Dai_fidelity_17}, 
mixed states of open systems \cite{Kshetrimayum_diss_17}, 
or exited states \cite{Vanderstraeten_tangentPEPS_15}.

Alongside iPEPS,
progress was made in simulating cylinders of finite width with DMRG. 
They are routinely used to investigate 2D ground states \cite{Simons_Hubb_17} and 
recently were applied also to 2D thermal states \cite{Stoudenmire_2DMETTS_17,Weichselbaum_Tdec_18}.
Among alternative approaches are methods of direct contraction and renormalization of
a 3D tensor network representing a 2D thermal density matrix \cite{Li_LTRG_11,Xie_HOSRG_12,Ran_ODTNS_12,Ran_NCD_13,Ran_THAFstar_18,Su_THAFoctakagome_17,Su_THAFkagome_17, Ran_Tembedding_18} and 
multi-scale entanglement renormalization ansatz (MERA) \cite{Vidal_MERA_07,Vidal_MERA_08,Evenbly_branchMERA_14,Evenbly_branchMERAarea_14}.

\section{Outline}
\label{sec:outline}

In this work we implement an algorithm to simulate real, Lindbladian and imaginary time evolution with iPEPS. 
The evolution operator is decomposed into small time steps using a Suzuki-Trotter decomposition \cite{Trotter_59,Suzuki_66,Suzuki_76}. 
Each time step is a product of 2-site nearest-neighbor gates.
It is applied to an iPEPS $|\psi\rangle$ with a bond dimension $D$. 
Every nearest-neighbor gate enlarges the dimension of the nearest-neighbor bond to which it is applied from $D$ to $k\times D$, with $k \leq d^2$ where $d$ is the local dimension of a lattice site. 
The new iPEPS represents a new state $|\psi'\rangle$. 
It is clear that repeated application of time steps would result in an exponential growth of the bond dimension. 
Therefore, 
after each time step
it is necessary to approximate the exact new iPEPS $|\psi'\rangle$ by an approximate iPEPS  -- 
representing a state $|\psi''\rangle$ --
with all bonds having the original bond dimension $D$. 
A straightforward optimization of the fidelity between the approximate $|\psi''\rangle$ and the exact $|\psi'\rangle$ is feasible \cite{OurExact}
and, in principle, it should give the most accurate $|\psi''\rangle$, but it is not the most efficient one.

In order to obtain a good approximation efficiently,
we consider an auxiliary iPEPS $|\tilde\psi''\rangle$ which is build from the same tensors as the new iPEPS $|\psi'\rangle$ 
except the two tensors at one of the bonds to which the nearest-neighbor gates were applied.
These two exact tensors --
each with the enlarged bond dimension $kD$ along this bond --
are replaced by two auxiliary tensors with the original bond dimension $D$.
These two auxiliary tensors are optimized to maximize fidelity between 
the exact $|\psi'\rangle$ and the auxiliary $|\tilde\psi''\rangle$.
Then the approximate $|\psi''\rangle$ is constructed by replacing {\it all} pairs of nearest-neighbor tensors by the optimal auxiliary tensors.

The optimization of the two auxiliary tensors in $|\tilde\psi''\rangle$ is much more efficient than optimization of 
an infinite number of copies of the two tensors in $|\psi''\rangle$ that is necessary to find the best $|\psi''\rangle$ 
straightforwardly \cite{OurExact}. Therefore, maximization of the fidelity between $|\psi'\rangle$ and $|\tilde\psi''\rangle$ (instead of $|\psi''\rangle$) is crucial for the efficiency. 
This local optimization is done in the exact environment of the new $|\psi'\rangle$ 
to give the best accuracy of the approximate $|\psi''\rangle$  
while still solving a {\it local} variational problem.

Furthermore, in order to make the best use of the limited $D$,
in the case of imaginary time evolution of a thermal state purification, 
we test applying disentangling unitary nearest-neighbor gates (disentanglers) to the ancillas at the same sites as the evolution nearest-neighbor gates. 
The disentanglers act on the ancilla indices and are optimized to minimize the necessary bond dimension $D$. 
This technique was used before for 1D MPS simulations \cite{dissentangler_quench,dissentangler_pollmann}.
Here for the first time it is implemented for 2D iPEPS where the bond dimension $D$ is a much more limited resource.

In the same case of thermal states,  
we also test an even more efficient optimization scheme. 
It is equivalent to the full update scheme that was used before
in imaginary time evolution of a pure state towards a ground state \cite{Cirac_iPEPS_08,fu}. 
In this scheme the original $|\psi\rangle$ -- 
with the smaller original D on all bonds --
is used as an environment for the optimized auxilliary tensors.
We benchmark this approximation for thermal states of the 2D quantum Ising model and find that it yields similar results as the ones obtained with  the exact environment of $|\psi'\rangle$.

A challenging application of the algorithm is real time evolution after a sudden quench of a parameter in a Hamiltonian. 
The quench excites entangled pairs of quasi-particles with opposite quasi-momenta that run away from each other and make 
the entropy of entanglement grow asymptotically linearly in time. 
Therefore, a tensor network is doomed to fail after a certain finite evolution time. 
Nevertheless, 
for 1D systems MPS proved to be a useful tool to simulate time evolution after sudden quenches, 
see e.g. \onlinecite{Zaunerstauber_DPT_17}. 
In this work we simulate a sudden quench of the transverse field
in the quantum Ising model to demonstrate that the same can be attempted with an iPEPS in 2D.
This opens prospects for simulation of many-body localization in 2D\cite{Wahl_MBL_17} 
for which excitations remain localized and the entanglement growth is slow.

The growth of the entanglement can also be slowed down, or even halted, by coupling to local Markovian environment \cite{Montangero_master_16,Kshetrimayum_diss_17}. 
We provide a proof of principle simulation of Lindbladian evolution for the 2D transverse field quantum Ising model subject to dissipation \cite{Kshetrimayum_diss_17}.

Last but not least, imaginary time evolution generating thermal states of a quantum Hamiltonian can be simulated efficiently. 
Both the thermal states of local Hamiltonians and iPEPS representations of thermal density operators obey the area law for mutual 
information making an iPEPS a promising ansatz for thermal states \cite{Wolf_Tarealaw_08}. 
In this paper we simulate thermal states of the 2D quantum Ising model in the vicinity of the second order phase transition. 
We add small symmetry breaking bias field $h_z$ in order to smooth the evolution across the critical point.
The smoothed evolution can be simulated accurately with modest computational resources. 
By extrapolation to $h_z=0$ we obtain accurate estimates of the critical temperature
even when the transverse field $h_x$ is close to the quantum critical point.

The paper is organized as follows.
In section \ref{sec:algorithm} we present an algorithm for the most general case of the time evolution of thermal state's purification with disentanglers. In section \ref{sec:results} we provide benchmark results for the 2D quantum Ising model. 
In subsection \ref{sec:resreal} we present results for real-time evolution after a global quench. Subsection \ref{sec:xxz} presents results for evolution with Markovian master equation, while subsections \ref{sec:resimag}, \ref{sec:resest} present results for thermal states. In subsections  \ref{sec:dis} and  \ref{sec:FU} we benchmark the algorithm with disentanglers and the full-update algorithm, respectively. In subsection \ref{sec:SU} we compare the algorithm for thermal states with a simple update algorithm. We conclude in section \ref{sec:conclusion}. Furthermore, in appendices \ref{app:reduced} and \ref{app:num} we provide technical details of the algorithm and the benchmark simulations.

\section{Exact-environment full update with disentanglers}
\label{sec:algorithm}
Here we introduce the algorithm in the most general case of simulation of thermal state purification 
by imaginary time evolution. We call the general algorithm an exact-environment (ee) full update (FU) 
with disentanglers (d) or eeFUd for short. Later we also consider   simplified
versions of the algorithm, in particular  a version  without  disentanglers (eeFU) and a version corresponding to the standard full update of  iPEPS tensors\cite{Cirac_iPEPS_08,Corboz_fiPEPS_10}  (FU), which is 
commonly used for ground state optimization. The algorithm for 
real time evolution of a pure state is obtained by ignoring ancillas alongside with the disentanglers 
applied to the ancillas. 

Our presentation of the general eeFUd algorithm is tailored for the quantum Ising model on an infinite square lattice 
that is going to be its testing ground in this paper:
\be 
H ~=~
- \sum_{\langle j,j'\rangle}Z_jZ_{j'}
- \sum_j \left( h_x X_j + h_z Z_j \right).
\label{calH}
\ee
Here $X,Z$ are the Pauli operators. At zero longitudinal field $h_z=0$ and zero temperature the model has 
a ferromagnetic phase with non-zero spontaneous magnetization $\langle Z \rangle$ for small enough 
transverse field $h_x$. The quantum critical point is $h_x=3.04438(2)\equiv h_0$\cite{Deng_QIshc_02}. For $h_x < h_0$ 
the model has a second order phase transition at finite temperature belonging to the 2D  classical Ising 
universality class. For $h_x=0$ it becomes the 2D classical Ising model with $T_c=2/\ln(1+\sqrt{2})\approx 2.269$.  

\subsection{Purification of thermal states}
\label{sec:purification}

\begin{figure}[t!]
\vspace{-0cm}
\includegraphics[width=0.9999\columnwidth,clip=true]{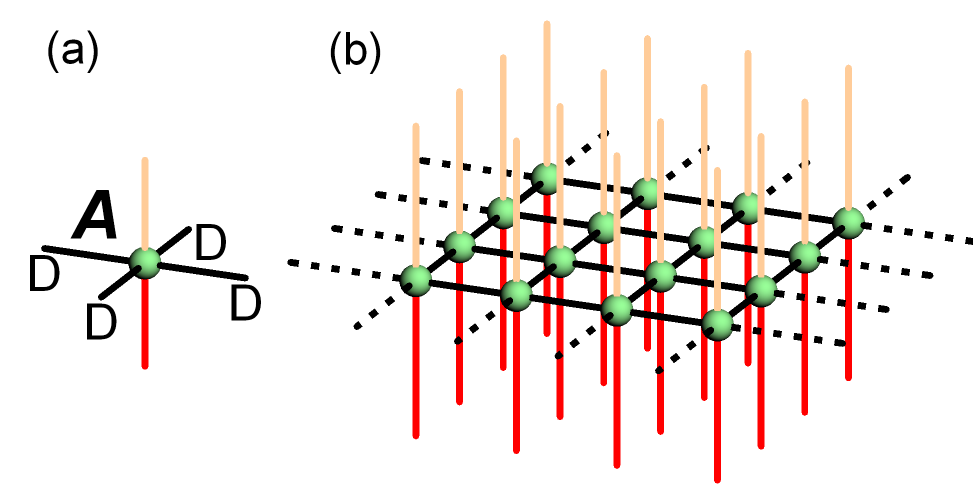}
\vspace{-0cm}
\caption{
In (a)
an elementary rank-6 tensor $A$ of a purification is shown. 
The top (orange) index numbers ancilla states $a=0,1$,
the bottom (red) index numbers spins states $s=0,1$,
the four (black) bond indices have a bond dimension $D$.
In (b)
an iPEPS representation of the purification.
Here pairs of elementary tensors at nearest-neighbor sites are contracted through their connecting bond indices.
The dotted lines connect with the rest of the infinite lattice. 
We obtain the iPEPS representation of a pure state by
reducing the dimension of ancilla indices to 1 or, equivalently, 
erasing the ancilla lines altogether.   
}
\label{fig:A}
\end{figure}

The structure of the algorithm is the most general in case of imaginary time evolution generating thermal states.
Their purifications are pure states in an enlarged Hilbert space of physical spins and virtual ancillas. 
Every spin with states $s=0,1$ is accompanied by an ancilla with states $a=0,1$. 
The space is spanned by states $\prod_j |s_j,a_j\rangle$, where $j$ numbers lattice sites. 
The Gibbs operator at an inverse temperature $\beta$ is obtained from its purification $|\psi(\beta)\rangle$ by tracing out the ancillas:
\be
\rho(\beta) \propto
e^{-\beta H} =
{\rm Tr}_a |\psi(\beta)\rangle\langle\psi(\beta)|.
\label{rhobeta}
\ee
At $\beta=0$ we choose the purification as a product over lattice sites,
\be
|\psi(0)\rangle = \prod_j ~\sum_{s=0,1} |s_j,s_j\rangle,
\label{psi0}
\ee
to initialize its imaginary time evolution
\be
|\psi(\beta)\rangle~=~
G(\beta) e^{-\frac12\beta H}|\psi(0)\rangle~=~
G(\beta) U(-i\beta/2)|\psi(0)\rangle.
\label{psibeta}
\ee
Here the evolution operator $U(\tau)\equiv e^{-i\tau H}$ acts in the Hilbert space of spins and 
$G(\beta)$ is an arbitrary unitary gauge transformation acting on ancillas.  
With the initial state (\ref{psi0}), equation (\ref{rhobeta}) becomes
\be
\rho(\beta) ~\propto~ U(-i\beta/2) U^\dag(-i\beta/2)
\label{UU}
\ee
with the gauge transformation cancelled out. 

Just like a pure state of spins, the purification can be represented by an iPEPS, see Fig.~\ref{fig:A}.
In the following we use the gauge freedom to minimize its bond dimension $D$.
Therefore $G(\beta)$ will often be referred to as a disentangler.

\subsection{Suzuki-Trotter decomposition}
\label{sec:ST}

In the second-order Suzuki-Trotter decomposition \cite{Trotter_59,Suzuki_66,Suzuki_76} the evolution operator is split into a product of small time steps, $U(\tau)=U(d\tau)^N$, and each small time step is approximated as
\bea
U(d\tau)
&=&
U_{h} (d\tau/2)
U_{ZZ}(d\tau  )
U_{h} (d\tau/2),
\label{Udbeta}
\eea
where 
\be
U_{ZZ}(d\tau) =
\prod_{\langle j,j'\rangle}e^{i d\tau Z_jZ_{j'}},~
U_{h}(d\tau)  =
\prod_j e^{i d\tau h_j}
\label{UZZ}
\ee
are elementary classical gates and $h_j = h_x X_j + h_z Z_j$. 
The action of the local gate $U_h$ on iPEPS is trivial:
it modifies every iPEPS tensor simply by acting on its 
physical index with $e^{i d\tau h_j}$.

\subsection{Sublattices}
\label{sec:sub}
In order to implement the gate $U_{ZZ}(d\tau)$, 
we divide the infinite square lattice into two sublattices A and B, with two different PEPS tensors at each sublattice,
see Fig.~\ref{fig:2site}(a).
On the A-B checkerboard the gate becomes a product of 4 commuting nearest-neighbor gates:
\bea
&& 
U_{ZZ}(d\tau) = U^x_0(d\tau) U^x_1(d\tau) U^y_0(d\tau) U^y_1(d\tau).
\label{UZZ2s}
\eea
Here $x$ ($y$) is the horizontal (vertical) direction spanned by $\vec{e}_x$ ($\vec{e}_y$),
\bea
U^x_s(d\tau)  &=&  \prod_{mn} e^{i d\tau Z_{2m+s-1,n}Z_{2m+s,n}},\label{Ua}\\
U^y_s(d\tau)  &=&  \prod_{mn} e^{i d\tau Z_{m,2n+s-1}Z_{m,2n+s}},\label{Ub}
\eea
and $Z_{m,n}$ is an operator at site $m\vec{e}_x+n\vec{e}_y$. 
For the sake of definiteness, 
in the following we focus on $U^x_0(d\tau)$.
The other nearest-neighbor gates are implemented analogously.

\subsection{Nearest-neighbour gate}
\label{sec:NNgate}
In order to facilitate application of $U^x_0(d\tau)$ to iPEPS, 
first of all we use a singular value decomposition to rewrite 
a 2-site term $e^{id\tau Z_jZ_{j'}}$ acting on a nearest-neighbor bond 
as a contraction of 2 smaller tensors acting on each site: 
\bea
e^{id\tau Z_jZ_{j'}} &=&
\sum_{\mu=0,1}
z_{j,\mu}
z_{j',\mu}.
\label{svdgate}
\eea
Here $\mu$ is a bond index with a bond dimension $2$ and 
$z_{j,\mu}\equiv\sqrt{\Lambda_\mu}\,(Z_j)^\mu$, 
where $\Lambda_0=\cos d\tau$ and $\Lambda_1=i\sin d\tau$.

\begin{figure}[t!]
\vspace{-0cm}
\includegraphics[width=0.9999\columnwidth,clip=true]{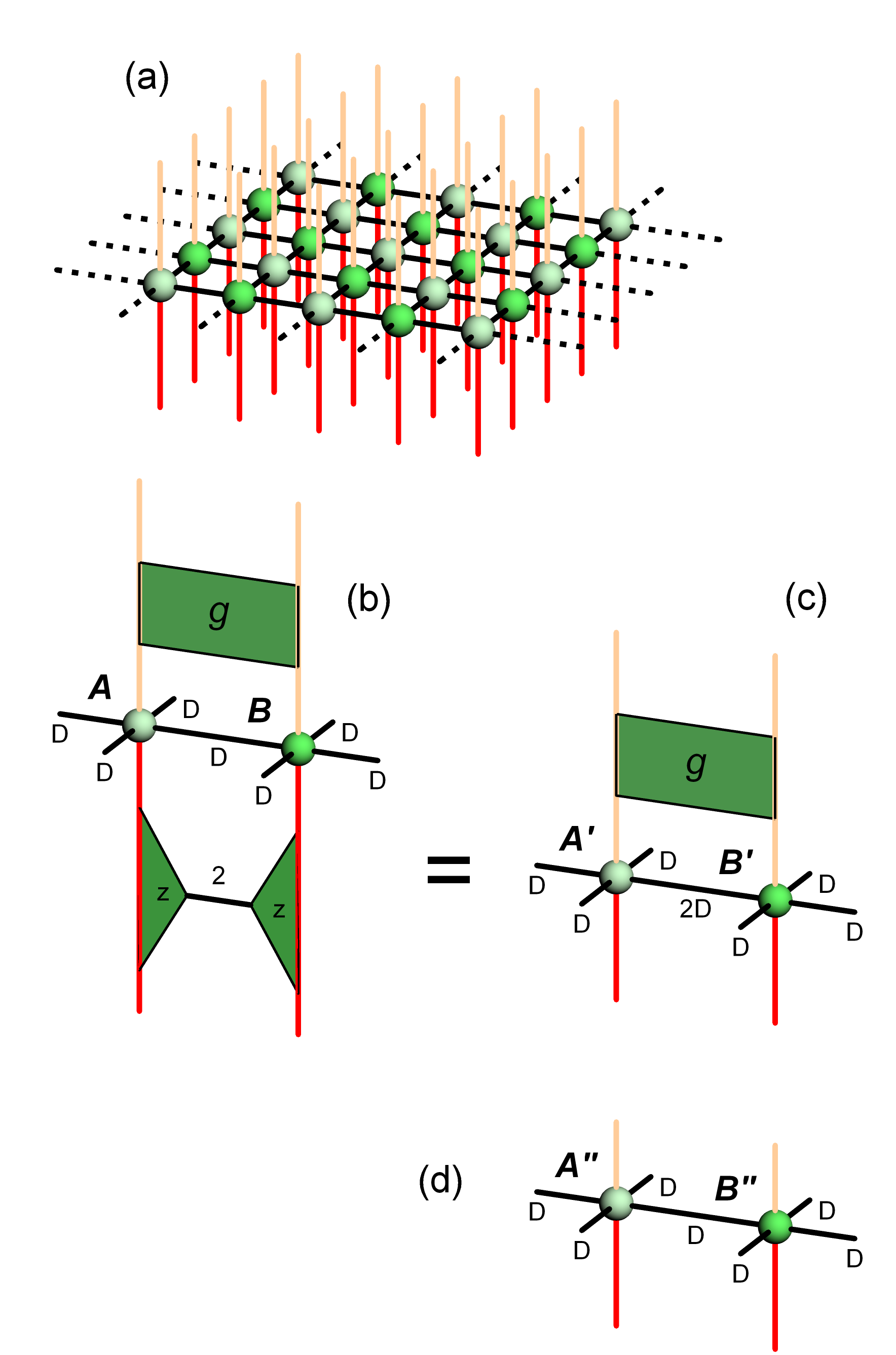}
\vspace{-0cm}
\caption{
In (a)
an infinite square lattice is divided into two sublattices with tensors $A$ (brighter green) and $B$ (green). 
In (b)
SVD decomposition of the nearest-neighbor gate in Eq. (\ref{svdgate}) is applied to spin indices of every considered nearest-neighbor pair of tensors $A$ and $B$. At the same time, a disentangler $g$ is applied to their ancilla indices. The result is an exact purification $|\psi'\rangle$.
In (c)
the tensors $A$ and $B$ can be conveniently fused with their respective pair of $z$ tensors becoming
new tensors $A'$ and $B'$ with a doubled bond dimension $2D$.
In (d)
the algorithm optimizes $g$ together with new tensors $A''$ and $B''$ -- that have the original bond dimension $D$ -- so that an approximate new iPEPS $|\psi''\rangle$ made of $A''$ and $B''$ has a maximal fidelity to the exact $|\psi'\rangle$. 
}
\label{fig:2site}
\end{figure}

Consequently, 
when the global gate $U^x_0(d\tau)$ is applied to 
the checkerboard iPEPS $|\psi\rangle$ with tensors $A$ and $B$ in Fig.~\ref{fig:2site}(a), 
then every pair of tensors $A$ and $B$ at nearest-neighbor sites $(2m-1)\vec{e}_x+n\vec{e}_y$ and $2m\vec{e}_x+n\vec{e}_y$ is applied with the SVD-decomposed nearest-neighbor-gate, see Fig.~\ref{fig:2site}(b). 
At the same time, 
its ancilla indices are applied with a nearest-neighbor disentangling gauge transformation $g$. 
The result is an exact purification $|\psi'\rangle=\tilde g U^x_0(d\tau)|\psi\rangle$, 
where $\tilde g$ is a tensor product of all $g$'s applied to all the considered nearest-neighbor bonds. 
When tensors $A$ and $B$ are contacted with their respective $z$'s, 
they become, respectively, 
$A'$ and $B'$ connected by an index with a doubled bond dimension $2D$, 
see Fig.~\ref{fig:2site}(c). 
Finally $|\psi'\rangle$ is approximated by a new iPEPS $|\psi''\rangle$ with the original bond dimension $D$ at every bond, see Fig.~\ref{fig:2site}(d). Its tensors $A''$ and $B''$ are optimized together with the disentangler $g$ in order to maximize fidelity between the exact $|\psi'\rangle$ and the approximate $|\psi''\rangle$.

\begin{figure}[t!]
\vspace{-0cm}
\includegraphics[width=0.9999\columnwidth,clip=true]{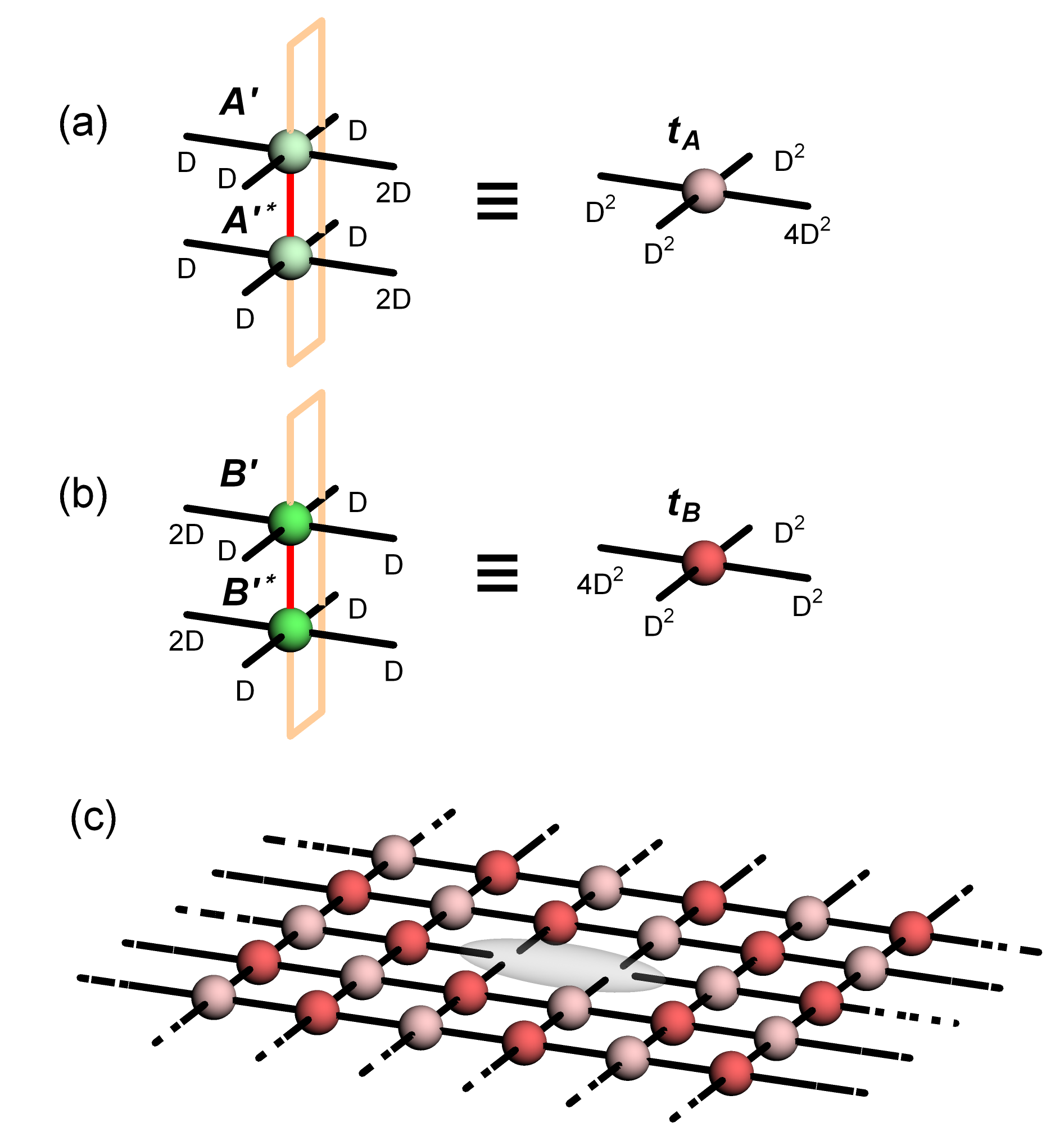}
\vspace{-0cm}
\caption{
In (a)
tensor $A'$ is contracted with its complex conjugate into a transfer tensor $t_A$.
In (b)
tensor $B'$ is contracted with its complex conjugate into a transfer tensor $t_B$.
In (c)
the tensor environment for the bond $A''-B''$.
The environment is a rank-6 tensor, each index has dimension $D^2$.
}
\label{fig:t}
\end{figure}

\subsection{Bond fidelity}
\label{sec:step}
The optimization aims at maximizing the global fidelity:
\be 
F=\frac{\langle\psi''|\psi'\rangle\langle\psi'|\psi''\rangle}{\langle\psi''|\psi''\rangle}.
\label{F}
\ee
A direct maximization of $F$ is feasible \cite{OurExact} but not the most efficient approach.

In order to introduce a more efficient algorithm,
we define an auxiliary state $|\tilde\psi''\rangle$, 
where the diagram in Fig.~\ref{fig:2site}(c) is replaced by Fig.~\ref{fig:2site}(d) 
not at all the considered bonds but only at one. 
In other words, at all the bonds the tensor network  $|\tilde\psi''\rangle$  
is the same as the exact  $|\psi'\rangle$ except at one particular bond.

The efficient algorithm maximizes local bond fidelity
\be 
\tilde F=
\frac{\langle\tilde\psi''|\psi'\rangle
      \langle\psi'|\tilde\psi''\rangle}
     {\langle\tilde\psi''|\tilde\psi''\rangle}
\label{tildeF}
\ee
with respect to $A''$, $B''$, and $g$. Once converged, $A''$ and $B''$ are placed at all sites in the new iPEPS $|\psi''\rangle$. 
This global placement of the locally optimized tensors is an approximation when compared to the global optimization in Ref. \onlinecite{OurExact}.

However, as the optimized bond in (\ref{tildeF}) is surrounded by the exact environment of $|\psi'\rangle$, then -- for $D$ large enough 
to cause negligible truncation errors --   the approximation should not compromise the accuracy in a significant way. 

The rank-6 bond environment in Fig.~\ref{fig:t}(c) is obtained approximately with the corner transfer matrix renormalization group (CTMRG) \cite{Baxter_CTM_78,Nishino_CTMRG_96,Orus_CTM_09,Corboz_CTM_14}. It is an approximate numerical method with a refinement parameter: an 
environmental bond dimension $\chi$. All following results were checked for convergence with increasing $\chi$. Considering numerical cost, 
CTMRG is the bottleneck of the algorithm. 

The unitary disentanglers accelerate the convergence with $D$ at no extra leading cost in the bottleneck CTMRG, because in all overlaps in Eq. (\ref{tildeF}) the disentanglers in bra and ket layers cancel out ($gg^\dag=1$) on all bonds except the optimized one. 

Therefore, the key to the efficiency is that in all the overlaps  in Eq. (\ref{tildeF}) the tensor environment 
for the optimized bond is the same, see Fig.~\ref{fig:t}(c), and depends neither on the optimized tensors $A''$ 
and $B''$ nor the disentangler $g$. It is, therefore, calculated only once.

\begin{figure}[t!]
\vspace{-0cm}
\includegraphics[width=0.7\columnwidth,clip=true]{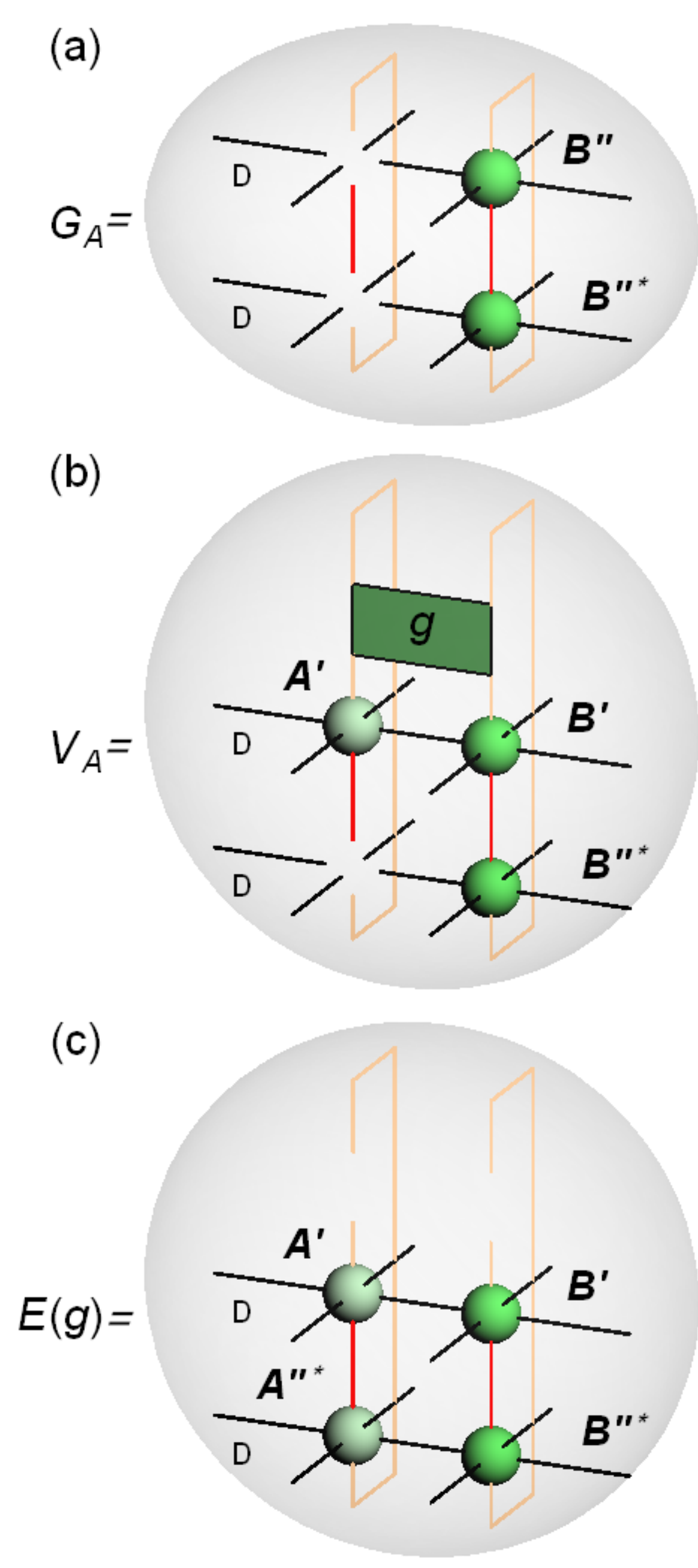}
\vspace{-0cm}
\caption{
In all panels,
each of the 6 indices of the environment in Fig.~\ref{fig:t}(c) was split into
a ket (upper) and a bra (lower) index, each of dimension $D$. 
In (a)
metric tensor $G_A$.
In (b)
gradient $V_A$.
In (c)
tensor environment for the disentangler $E(g)$.
}
\label{fig:GVE}
\end{figure}

\subsection{Optimization loop}
\label{sec:loop}

Tensors $A''$ and $B''$ and the disentangler $g$ are optimized iteratively in a loop until $\tilde F$ is maximized. Up to normalization, the best $|\tilde\psi''\rangle$ that maximizes $\tilde F$ also minimizes the norm
\bea
&&
\left|\left| |\tilde\psi''\rangle - |\psi'\rangle \right|\right|^2 =\nonumber\\
&&
\langle \tilde\psi''|\tilde\psi'' \rangle - 
\langle \tilde\psi''|\psi' \rangle -
\langle \psi'|\tilde\psi'' \rangle + 
\langle \psi'|\psi' \rangle.
\label{norm}
\eea
For fixed $A''$ and $B''$, the norm is linear in $g$. 
Therefore, when we define a tensor environment for $g$:
\be 
E(g)=\frac{\partial \langle\psi'|\tilde\psi''\rangle}{\partial g^* },
\ee
see Fig.~\ref{fig:GVE}(c), the norm is minimized by $g=uv^\dag$, where the unitary $u$ and $v$ come from a singular value decomposition $E(g)=u\lambda v^\dag$.

For fixed $B''$($A''$) and $g$, the norm (\ref{norm}) is quadratic in $A''$($B''$).
Therefore, when we define metric tensors and gradients:
\bea
G_A&=&
\frac{\partial^2\langle\tilde\psi''|\tilde\psi''\rangle}
     { \partial \left(A''\right)^* \partial \left(A''\right) },~~
V_A=
\frac{\partial \langle\tilde\psi''|\psi'\rangle}{\partial \left(A''\right)^* },
\label{GVA}\\
G_B&=&
\frac{\partial^2\langle\tilde\psi''|\tilde\psi''\rangle}
     { \partial \left(B''\right)^* \partial \left(B''\right) },~~
V_B=
\frac{\partial \langle\tilde\psi''|\psi'\rangle}
     {\partial \left(B''\right)^* },
\label{GVB}
\eea
see Figs. \ref{fig:GVE}(a,b),
the quadratic form is minimized by $A''=G_A^{-1}V_A$ $\left(B''=G_B^{-1}V_B\right)$,
where in practice the inverse means a pseudoinverse.

The optimizations are repeated in a loop 
\be
\dots\to g \to A''\to B'' \to \dots
\label{loop}
\ee
until a self-consistency is achieved and $\tilde F$ is converged.

Finally, to reduce the numerical cost of the algorithm in actual calculations we do not work with the full tensors $A''$ and $B''$, but 
with smaller reduced tensors $A''_{\rm red}$ and $B''_{\rm red}$ 
described in App. \ref{app:reduced} . 

\subsection{Full update imaginary time evolution}
\label{sec:fu}

As explained before, the environmental CTMRG procedure is the numerical bottleneck. 
In the presented general eeFUd algorithm, 
the environment is the exact $|\psi'\rangle$ with the enlarged bond dimension $2D$ (or $kD$ in general) on the considered bonds.
It is a natural question if the $|\psi'\rangle$-environment could be replaced by a more efficient $|\psi\rangle$-environment with 
$D$ on all bonds. This would reduce the algorithm to its simplified FUd version.

At first glance, the answer is no. 
As $|\psi\rangle$ differs from the exact $|\psi'\rangle$ by an error linear in $d\beta$,  
then also the environment differs from the exact one by an error linear in $d\beta$. 
In a simple model of error propagation --  
assuming that the environment error causes an error of $|\psi''\rangle$ 
proportional to the error of the environment, i.e., linear in $d\beta$
and that an error of the final state is a sum of errors at all intermediate steps --
the error of the final state does not depend on $d\beta$ and, therefore, 
it cannot be eliminated by decreasing $d\beta$.   

However, below in section \ref{sec:FU} we present numerical evidence that -- at least for evolution across a thermal critical point that is smoothed by a tiny symmetry breaking bias -- the approximate environment makes negligible difference to the results. We demonstrate also that results extrapolated to the zero bias limit are mutually consistent. 

\subsection{Real time evolution}
\label{sec:restep}

In addition to missing ancilla lines and disentanglers, 
one more simplification occurs in case of real time evolution of pure states.
As the nearest-neighbor gates are unitary, 
they cancel out in the overlaps in Eq. (\ref{tildeF}) at all bonds except the optimized one with $A''$ and $B''$.
Consequently,
in Figs. \ref{fig:t}(a,b) the tensors $A'$ and $B'$ can be substituted by $A$ and $B$.
All indices of the transfer tensors, $t_A$ and $t_B$, have the same dimension $D^2$
speeding up the bottleneck CTMRG procedure. 
Therefore,  for real-time evolution, the eeFU algorithm
simplifies to the FU algorithm.

\section{Results}
\label{sec:results}

\begin{figure}[t!]
\vspace{-0cm}
\includegraphics[width=0.9999\columnwidth,clip=true]{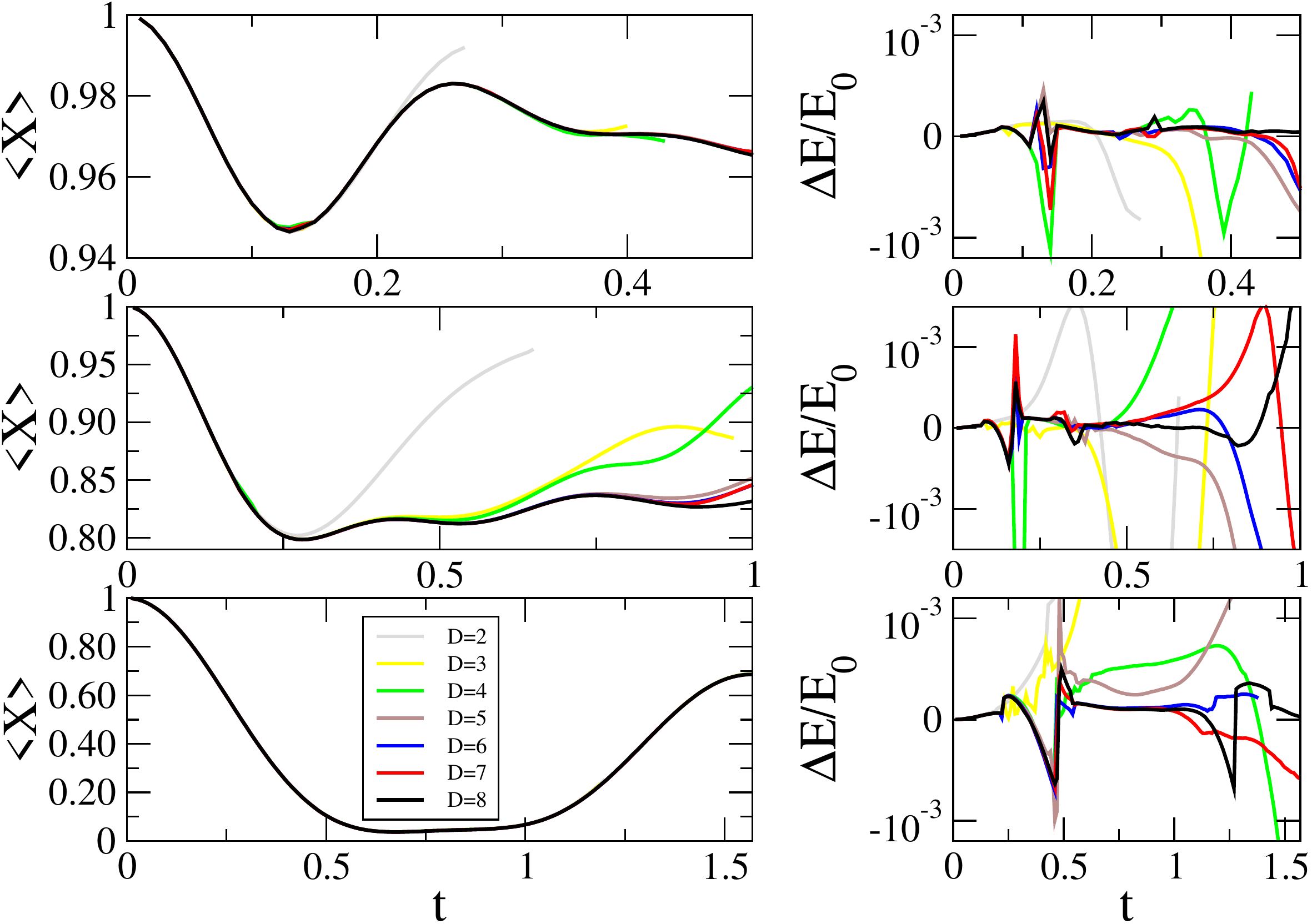}
\vspace{-0cm}
\caption{
Transverse magnetization $\langle X \rangle$ (left column) and relative energy error per site
(right column) after a sudden quench from a ground state
in a limit of $h_x\to\infty$, with all spins pointing along $x$, 
down to a finite $h_x=2h_0$ (top row), $h_x=h_0$ (middle row), and $h_x=h_0/10$ 
(bottom row). 
With increasing bond dimension $D=2,...,8$ the magnetization appears converged for increasingly long times.
The relative energy error is defined as $\Delta E/E_0$, where $E_0=-h_x$ is
the initial energy at $t=0^+$ right after the quench and $\Delta E=E(t)-E_0$
is the energy error.
The energy conservation shows improvement with increasing $D$. 
}
\label{fig:quench}
\end{figure}

\subsection{Real time evolution after quench}
\label{sec:resreal}

We begin with simulations of a real time evolution in the Ising model (\ref{calH}) without the bias field, $h_z=0$.
A sudden quench is considered from a limit of $h_x\to\infty$ down to $h_x=2h_0$ within the same paramagnetic phase, 
$h_x=h_0$ at the quantum critical point, and $h_x=h_0/10$ deep in the ferromagnetic phase.
The initial state is the ground state with all spins pointing along $x$ and 
full transverse magnetization $\langle X \rangle=1$.
Figure \ref{fig:quench} shows the magnetization $\langle X\rangle$ and relative energy error per site $E$ 
after the sudden quench at $t=0$ for different bond dimensions. 
With increasing $D$ the overall energy conservation improves and the transverse magnetization appears 
converged for longer times.

The simulations converge the fastest for $h_x=h_0/10$. This transverse field is close to $h_x=0$
when the Hamiltonian becomes classical and $D=2$ is enough to represent the evolution exactly.
More physically,
at $h_x=0$ quasiparticles have a flat dispersion relation, 
hence they are not able to propagate and spread entanglement.
The opposite case is the critical $h_x=h_0$ when the quasiparticles are gapless,
hence they are excited in large numbers and propagate quickly.
Here the convergence with increasing $D$ is slower.

\subsection{Markovian master equation}
\label{sec:xxz}

\begin{figure}[tb!]
\vspace{-0cm}
\includegraphics[width=0.9999\columnwidth,clip=true]{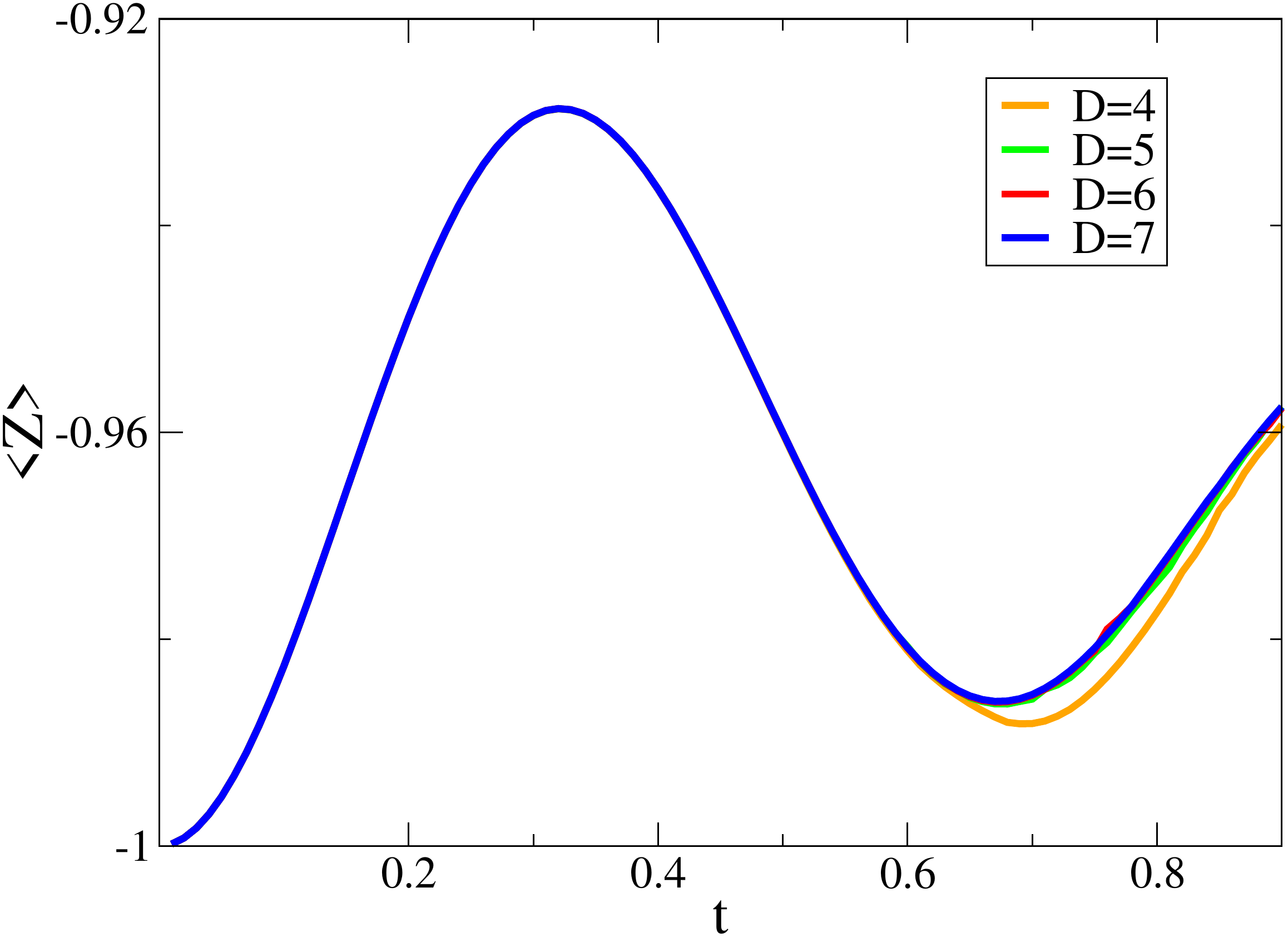}
\vspace{-0cm}
\caption{  
Time evolution of the longitudinal magnetization $\langle Z\rangle$ with the Markovian master equation
(\ref{me}) initialized with all spins pointing down: $\langle Z \rangle=-1$.
Different colors correspond to different bond dimensions.
With increasing $D$ the curves appear converged over an increasing range of
evolution time.}
\label{fig:me}
\end{figure}

After vectorization of a density matrix $\rho$, 
the real time algorithm can be easily adapted to evolve a Markovian master equation \cite{ZwolakVidal,Kshetrimayum_diss_17}. 
The vectorized $\rho_{\rm vec}$ is represented by an iPEPS that is isomorphic to an iPEPO representing the density matrix 
operator $\rho$.

We test the algorithm for the Lindblad master equation \cite{Kshetrimayum_diss_17}
\be 
\dot\rho=
-i[H,\rho]+
\sum_j \left( L_j\rho L_j^\dag -\frac12\left\{ L_j^\dag L_j , \rho \right\} \right).
\label{me}
\ee
Here the Hamiltonian is again the quantum Ising model on an infinite square lattice,
\be 
H=\frac{V}{4}\sum_{\langle j,j'\rangle}Z_jZ_{j'}+\frac{h_x}{2}\sum_j X_j,
\label{Hme}
\ee
and $L_j=\sqrt{\gamma}(X_j-iY_j)/2$ is a spin lowering operator. 

We set the dissipation rate $\gamma=1$ and,
as in Ref. \onlinecite{Kshetrimayum_diss_17},
consider the interaction strength $V=5\gamma$.
We choose $h_x=2\gamma$ and an initial state with all spins polarized down: $\langle Z_j\rangle=-1$. 
Figure \ref{fig:me} shows the longitudinal magnetization $\langle Z\rangle$ in function of time.
With increasing $D$ the magnetization plots appear to converge over an increasing range of time.

\subsection{Thermal states with the eeFU algorithm}
\label{sec:resimag}

\begin{figure}[tb!]
\vspace{-0cm}
\includegraphics[width=0.9999\columnwidth,clip=true]{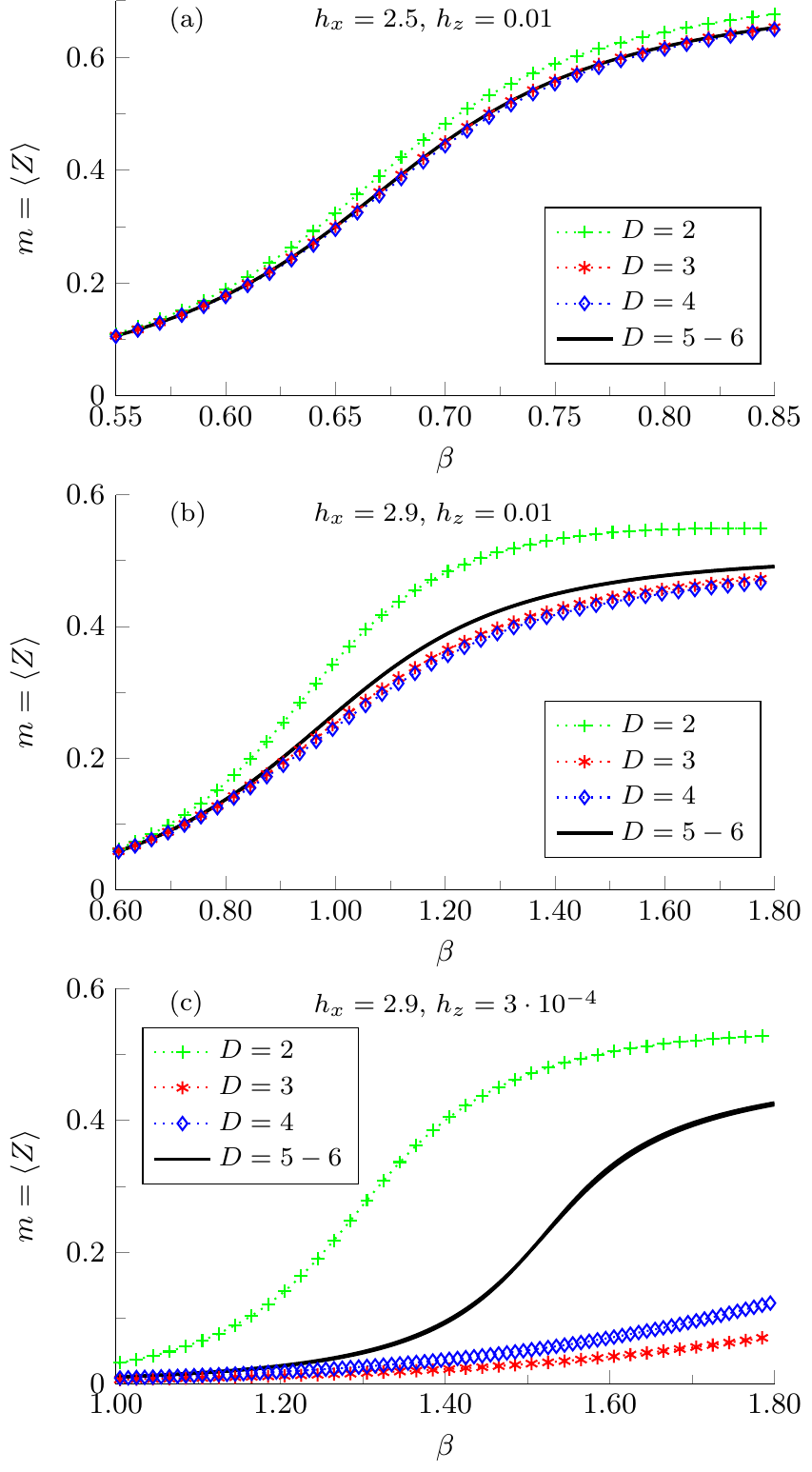}
\vspace{-0cm}
\caption{
Thermal states obtained by imaginary time eeFU evolution of a purification. In (a) and (b) the longitudinal magnetization $m=\langle Z \rangle$ as a function of the inverse temperature $\beta$ for longitudinal symmetry breaking bias field $h_z = 0.01$ and two different values of the  transverse field $h_x=2.5$ (a) and $h_x=2.9$ (b). With increasing $D$ the magnetization plots appear to converge. The $D=5$ magnetization plot appears to be  already converged.
The convergence is slower for the larger  transverse field $h_x=2.9$ for which quantum fluctuations are stronger.
In (c) we show $m$  for smaller  $h_z = 3\cdot10^{-4}$ and $h_x=2.9$.
Smaller $h_z$ makes simulations more demanding, but again the $D=5$ magnetization plot
 appears to be  converged.
}
  
\label{fig:imag}
\end{figure}

 \begin{figure}[tb!]
\vspace{-0cm}
\includegraphics[width=0.9\columnwidth,clip=true]{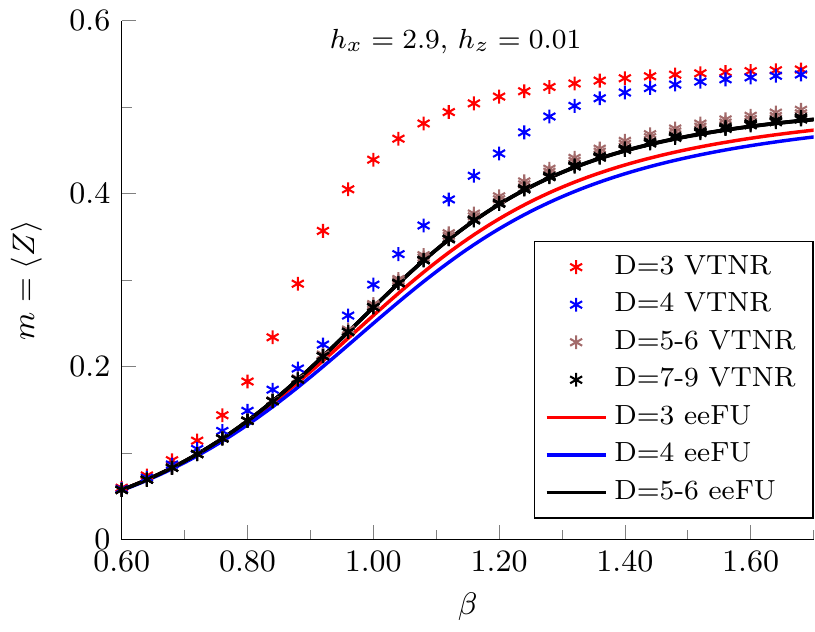}
\vspace{-0cm}
\caption{  
A comparison of the eeFU with variational tensor network renormalization (VTNR)\cite{ Czarnik_compass_16,Czarnik_VTNR_15,Czarnik_fVTNR_16} for $h_x=2.9$ and $h_z=0.01$. 
Both methods converge to each other with increasing $D$. 
The convergence is faster for the eeFU algorithm. 
}
\label{fig:evvsVTNR}
\end{figure}

In this subsection  
we analyze results obtained by the imaginary time evolution of a thermal state purification with the eeFU algorithm.
Here we neither use disentanglers nor the cheaper environment computed using  $|\psi\rangle$, that we postpone  to subsections \ref{sec:dis} and \ref{sec:FU}, respectively.
The eeFU results presented here will serve as a benchmark for the other methods.

We generate thermal states for transverse 
fields $h_x=2.5$ and $h_x=2.9$. Quantum Monte Carlo $T_c$ estimates
for these fields are $T_c=1.2737(6)$ and  $T_c=0.6085(8)$, respectively\cite{Hesselmann_TIsingQMC_16}. Both differ significantly 
from Onsager's $T_c \approx 2.269$ at $h_x=0$ demonstrating that for these fields quantum fluctuations are strong. 
Particularly challenging is the case of $h_x=2.9$ which is close to the quantum critical point  at 
$h_x=3.04438(2)$, see Ref. \onlinecite{Deng_QIshc_02}.    

We observe that close to the critical point, characterized by infinite correlation length, CTMRG convergence is very slow. Due to non-analytic $\beta$-dependence the results are also very sensitive to the time-step $d\beta$. Therefore, converging results near the critical point would be very expensive. In order to reduce these problems, we introduce a small longitudinal bias field $h_z$ which takes the state away from the critical one.

In Fig.~\ref{fig:imag} we show convergence with $D$ of the magnetization  $m(\beta)=\langle Z(\beta)\rangle$ obtained with  $0.0003 \le h_z \le 0.01$. In Fig.~\ref{fig:evvsVTNR} we compare the $m$ obtained by the eeFU method with that from the variational tensor network renormalization (VTNR) for $h_x=2.9$ and $h_z=0.01$ \cite{ Czarnik_compass_16,Czarnik_VTNR_15,Czarnik_fVTNR_16}. 
We see that both methods converge to each other, though the eeFU approach converges faster.

\subsection{Estimation of critical temperatures}
\label{sec:resest}

\begin{figure}[tb!]
\vspace{-0cm}
\includegraphics[width=0.9\columnwidth,clip=true]{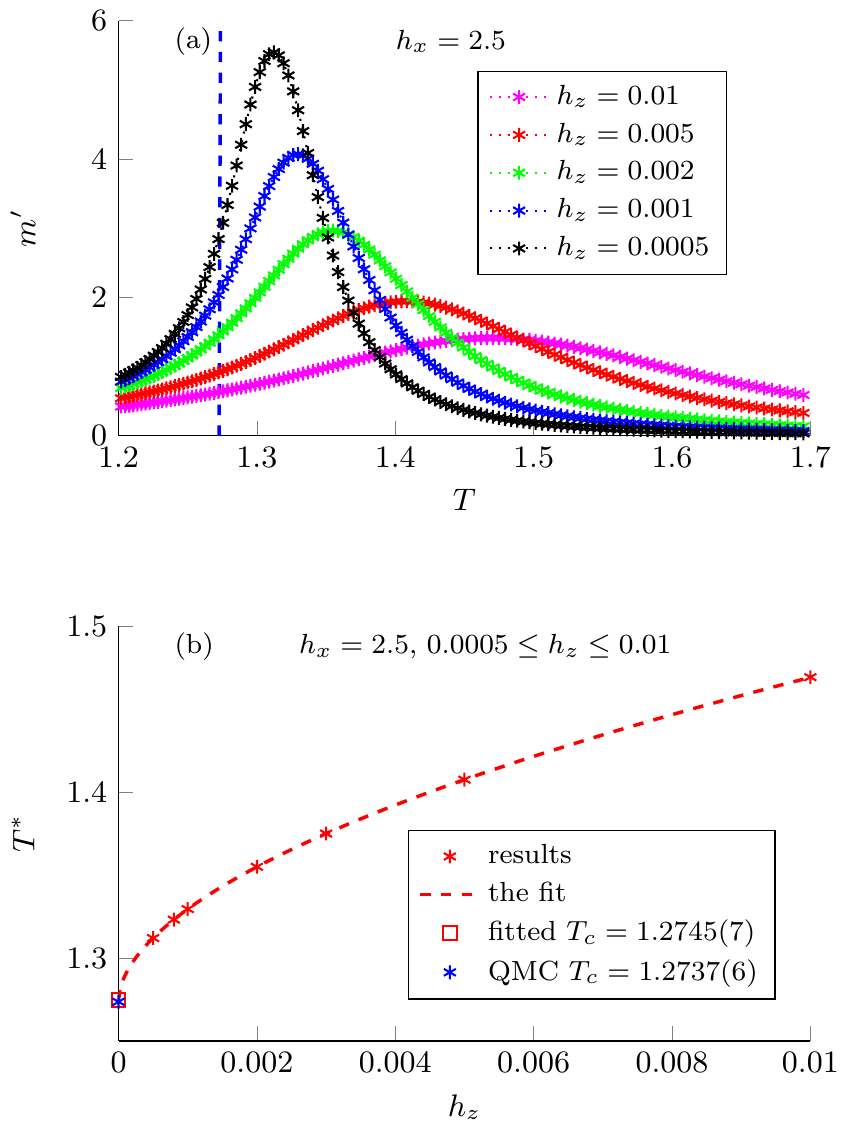}
\vspace{-0cm}
\caption{  
In (a) $m'(T)$  for $h_x=2.5$ and $D=5$. 
The blue dashed line shows the QMC  $T_c$. 
In (b)  $T_c$ is estimated by fitting the scaling (\ref{scalpeak}) of $m'$ peaks $T^{*}(h_z)$. 
Here we use  $0.0005 \le h_z \le 0.01$ and $D=5$. We obtain $T_c=1.2745(7)$ and  $1/\tilde\beta \delta =  0.549(4)$. 
The obtained $T_c$ agrees with the QMC  $T_c=1.2737(6)$ \cite{Hesselmann_TIsingQMC_16}, 
while obtained  $1/\tilde\beta\delta$ is close to the exact value $1/\tilde\beta\delta=8/15\approx 0.533$
(it differs from the exact value  by $3\%$).
}
\label{fig:Tcgx2p5peak}
\end{figure}

\begin{figure}[tb!]
\vspace{-0cm}
\includegraphics[width=0.9\columnwidth,clip=true]{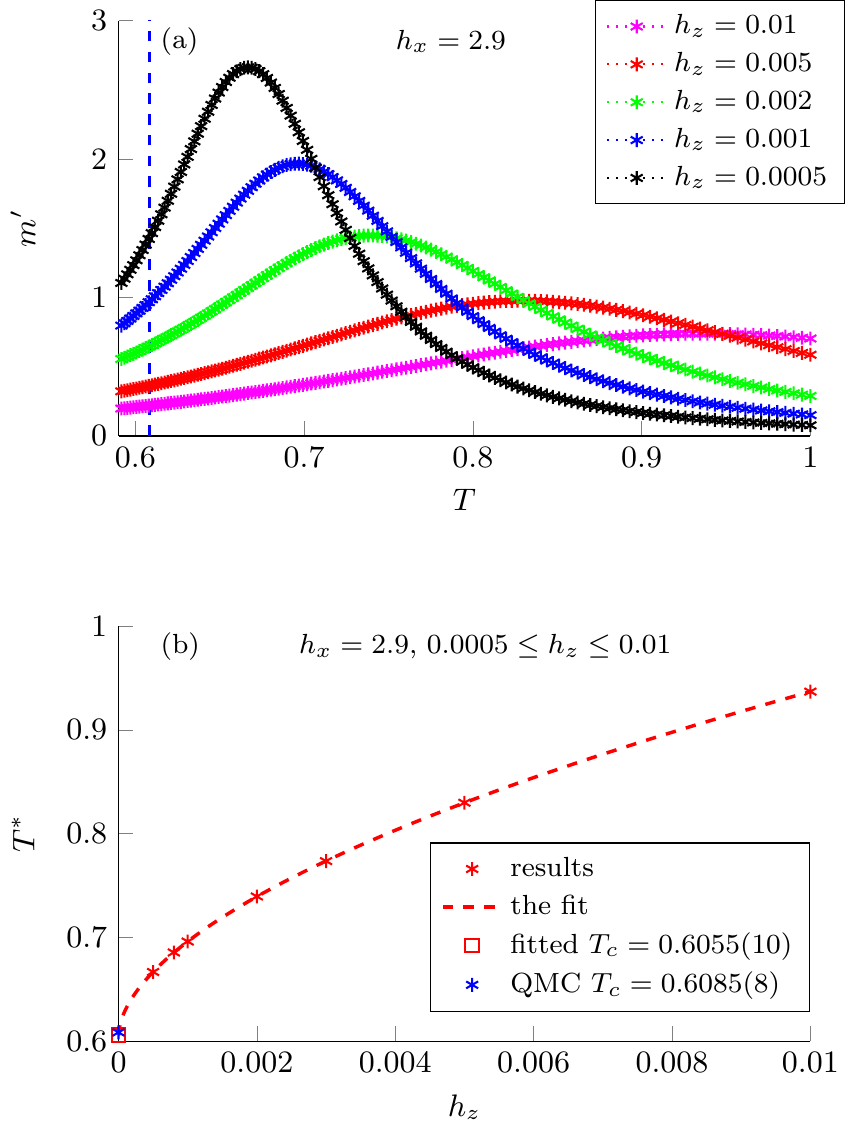}
\vspace{-0cm}
\caption{  
In (a) $m'(T)$  for $h_x=2.9$ and $D=5$. 
The blue dashed line shows the QMC  $T_c$. 
In (b)  $T_c$ and  $1/\tilde\beta\delta$ are estimated by fitting the scaling (\ref{scalpeak}) of $m'$ peaks $T^{*}(h_z)$. 
Here we use   $0.0005 \le h_z \le 0.01$ and $D=5$ obtaining $T_c = 0.6055(10)$ and $1/\tilde\beta \delta =  0.563(4)$. 
}
\label{fig:Tcgx2p9peak}
\end{figure}

In order to estimate the critical temperature $T_c$ of the second order phase transition, 
we assume that the order parameter $m$, that is converged in $D$,   
can be obtained for a symmetry breaking field $h_z$ that is small enough to fall into the scaling regime.  
This assumption leads to scaling ansatzes for $m$ and its derivative $m'$\footnote{We note that a scaling 
ansatz for $m'$ was proposed recently in a different context in Ref. \onlinecite{Corboz_FCLS_18}.} with 
respect to $t=(T-T_c)/T_c$:  
\bea
m  &\sim&  h_z^{1/\delta} f(t h_z^{-1/\tilde\beta\delta}),\\
m' &\sim&  h_z^{(\tilde\beta-1)/\tilde\beta\delta} g(t h_z^{-1/\tilde\beta\delta}).
\label{mderivscal}
\eea
Here $f$ and $g$ are non-universal scaling functions and $\delta$ and $\tilde{\beta}$ are 
critical exponents of the phase transition. We use here $\tilde{\beta}$
instead of conventional $\beta$ to distinguish it from the inverse temperature $\beta$. 
For finite $h_z$ the derivative $m'$ has a peak at temperature $T^{*}>T_c$. 
Therefore from (\ref{mderivscal}) follows scaling of $T^{*}$:
\be
T^{*} - T_c \sim  h_z^{1/\tilde\beta\delta}. 
\label{scalpeak}
\ee
We use this scaling to estimate $T_c$ and $1/\tilde\beta\delta$.

For $h_x=2.5$, 
using $0.0005 \le h_z \le 0.01$ and $D=5$
we obtain $T_c = 1.2745(7)$ and $1/\tilde\beta \delta =  0.549(4)$, see Fig.~\ref{fig:Tcgx2p5peak}. 
We find that the results are converged in the Suzuki-Trotter  step $d\beta$ and the environmental bond dimension $\chi$ 
for $d\beta = 0.002$ and $\chi = 25$, respectively. 
The fitted $T_c$ agrees with the QMC estimate $T_c=1.2737(6)$\cite{Hesselmann_TIsingQMC_16}  
and $1/\tilde\beta\delta$ is within $3\%$ of the exact value $8/15$. 

For $h_x = 2.9$ quantum fluctuations are stronger making the estimation more challenging. 
Using $0.0005 \le h_z \le 0.01$ and $D=5$
we obtain $T_c = 0.6055(10)$ and $1/\tilde\beta \delta =  0.563(4)$, see Fig.~\ref{fig:Tcgx2p9peak}. 
These estimates are within $0.5\%$ and $6\%$, respectively, of the QMC's $T_c=0.6085(8)$\cite{Hesselmann_TIsingQMC_16} 
and the exact $1/\tilde\beta \delta=8/15$. 
We find that results are converged in the Suzuki-Trotter step and the environmental bond dimension 
for $d\beta = 0.005$ and  $\chi = 25$, respectively.

\subsection{Thermal states with the eeFUd algorithm}
\label{sec:dis}

\begin{figure}[tb!]
\vspace{-0cm}
\includegraphics[width=0.99\columnwidth,clip=true]{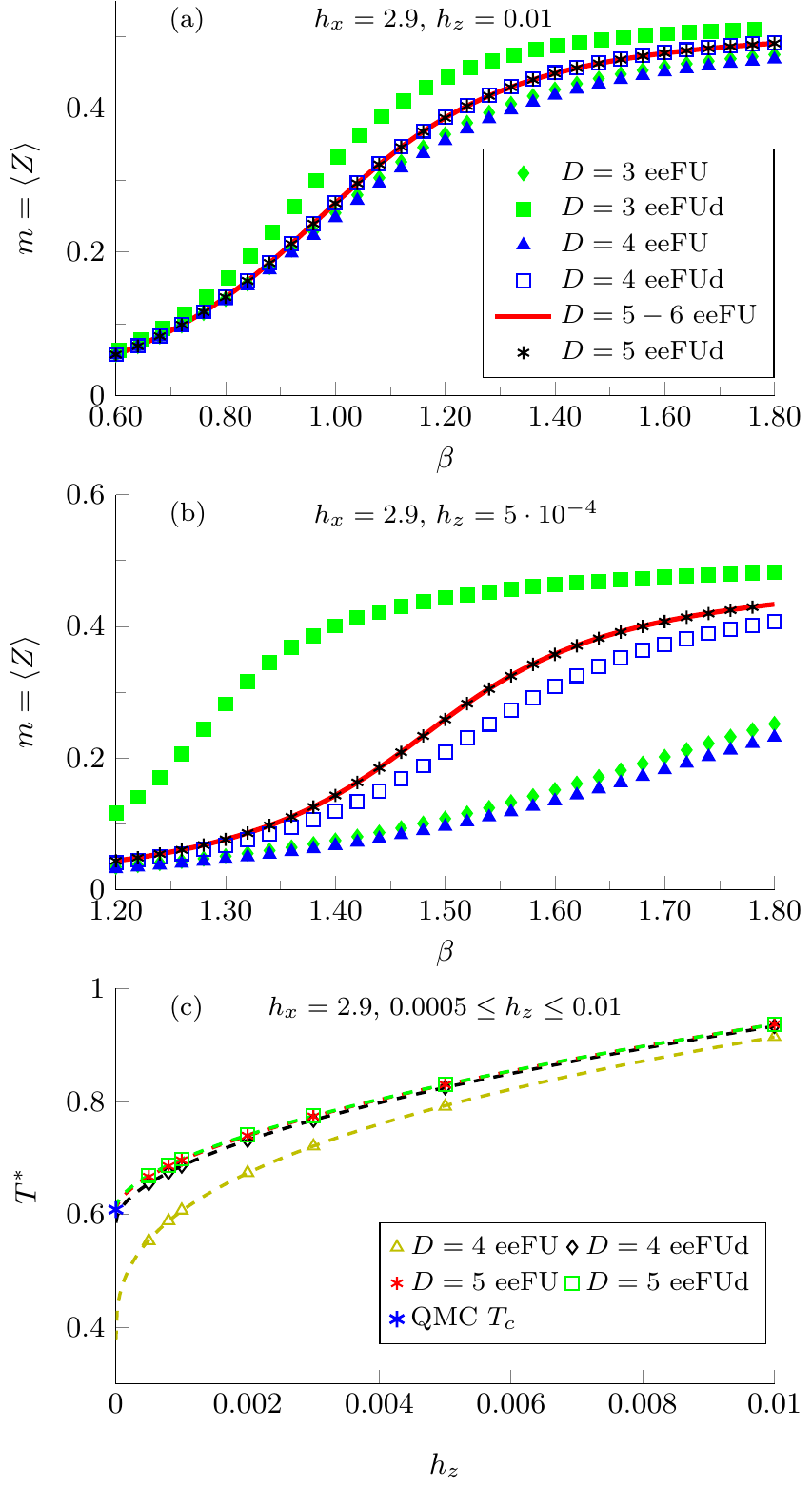}
\vspace{-0cm}
\caption{
In (a) and  (b) we show the longitudinal magnetization $m=<Z>$ in function 
of the inverse temperature $\beta$  obtained with (eeFUd) and without
(eeFU) the disentanglers for the transverse field $h_x=2.9$. 
In (a) results for the longitudinal symmetry breaking field $h_z=0.01$. In (b) results for 
more demanding $h_z=5\cdot 10^{-4}$. In both cases the $D=4$
 results with the disentanglers are much closer to the  converged  $D=5-6$  results.
 We see also that the $D=5$ results with the disentanglers are equivalent to the $D=5-6$ results obtained  without them.
 In (c)  $T_c$ and  and $1/\tilde\beta\delta$ 
estimation  by  the  scaling (\ref{scalpeak}) for $0.0005 \le h_z \le 0.01$ 
and $D=4,5$. The results corroborate the conclusions drawn from (a) and (b).
 Numerical values of the fitted $T_c$ and $1/\tilde\beta\delta$ are listed in Tab.~\ref{tab:dis}. 
}
\label{fig:dis}
\end{figure}

\begin{table}[tb!]
\begin{tabular}{|l|c|c|c|}
\hline
method & $D$ & $T_c$ & $1/\tilde\beta\delta$ \\ 
\hline
eeFU  & $4$ &  $0.38(2)$ & $0.37(3)$ \\
eeFUd & $4$ &  $0.582(2)$ & $0.539(5)$ \\
eeFU  & $5$ &  $0.6100(7)$ & $0.571(3)$ \\
eeFUd & $5$ &  $0.6099(8)$ & $0.569(3)$ \\
QMC\cite{Hesselmann_TIsingQMC_16} & - & $0.6085(8)$ & - \\
exact & - & - & $8/15 \approx 0.533$\\
\hline
\end{tabular}
\caption{Comparison of $T_c$ and $1/\tilde\beta\delta$ obtained in Fig.~\ref{fig:dis}(c) for $h_x=2.9$ using $0.0005 \le h_z \le 0.01$ with the eeFU and eeFUd algorithms. For $D=4$ the disentanglers improve accuracy by one order of magnitude. For $D=5$ both methods give the same accuracy. In order to enable direct comparison of the time evolution with and without the disentanglers we use here the same reduced tensors for both methods, see App. \ref{app:reduced}. The usage of the larger reduced tensors accounts for a small discrepancy between the $D=5$ eeFU result without the disentangler shown here and the eeFU result in Fig.~\ref{fig:Tcgx2p9peak} that was obtained with smaller reduced tensors.}
\label{tab:dis}
\end{table}


Next we test the eeFUd algorithm with disentanglers, comparing it to the eeFU algorithm without the disentanglers for the more challenging $h_x=2.9$. 
We compare the magnetization and $T_c, 1/\tilde\beta\delta$ obtained by the scaling (\ref{scalpeak}), 
see Fig.~\ref{fig:dis} and Tab. \ref{tab:dis}.  
We see that for $D=4$ results obtained with disentanglers are closer to convergence in $D$ than those without disentanglers. 
For $D=4$ the $T_c$ and $1/\tilde\beta\delta$ estimated with disentanglers are an order of magnitude more accurate than those without disentanglers.
For $D=5$, 
which is large enough to obtain good accuracy also without disentanglers, 
both methods give similar results. 

We conclude that it is possible to improve accuracy by using the disentanglers. The better accuracy with 
disentanglers comes at a price of more iterations of the optimization loop (\ref{loop}) and larger reduced 
tensors, see App.~\ref{app:reduced}. The cost of an iteration of the optimization loop is sub-leading as 
compared to the cost of the CTMRG.

\subsection{Thermal states with the FU algorithm}
\label{sec:FU}

\begin{figure}[tb!]
\includegraphics[width=0.9\columnwidth,clip=true]{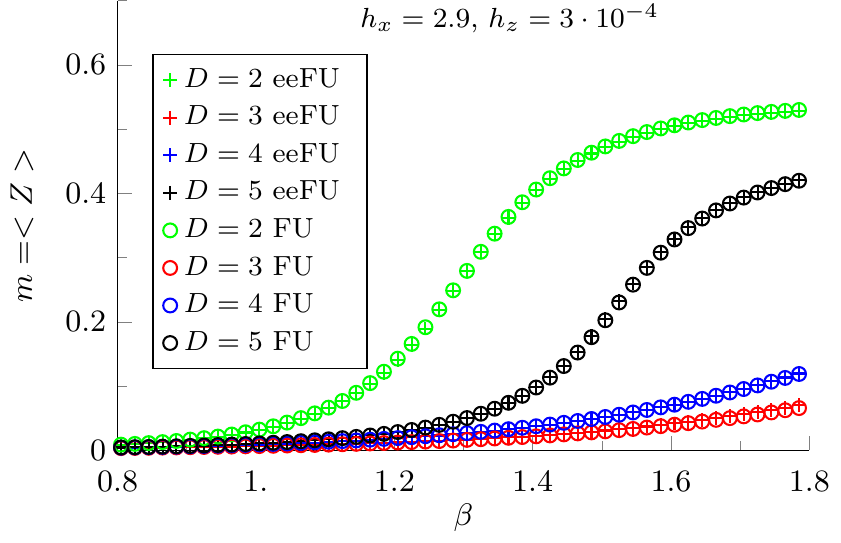}
\caption{
A comparison of thermal states obtained by the FU  and the eeFU algorithms. 
Here $h_x=2.9$ and $h_z=3\cdot 10^{-4}$. Both algorithms give similar results.  
}
\label{fig:FU}
\end{figure}
   
\begin{table}[tb!]
\begin{tabular}{|l|c|c|c|c|}
\hline
method  & $h_x$ & $D$ &  $T_c$ & $1/\tilde\beta\delta$ \\ 
\hline
eeFU   & $2.5$ & $5$ & $1.2745(7)$ & $0.549(4)$  \\
 FU   & $2.5$ & $5$ & $1.2746(6)$ & $0.549(3)$  \\
QMC\cite{Hesselmann_TIsingQMC_16} & $2.5$ & - & $1.2737(6)$ & - \\
eeFU   & $2.9$ & $5$ & $0.6055(10)$  & $0.563(4)$ \\
 FU   & $2.9$ & $5$ & $0.6061(19)$ & $0.564(7)$ \\
QMC\cite{Hesselmann_TIsingQMC_16} & $2.9$ & - & $0.6085(8)$ & - \\
exact &   -   &  -  &      -      & $8/15 \approx 0.533$\\
\hline
\end{tabular}
\caption{Comparison of $T_c$ and $1/\tilde\beta\delta$ obtained for $h_x=2.5$ and $h_x=2.9$  by the FU  and 
the eeFU algorithms. We use here the peak scaling (\ref{scalpeak}) for $0.0005 \le h_z \le 0.01$. 
Both algorithms have similar accuracy. }
\label{tab:FU}
\end{table}

We compare results obtained by the more efficient FU algorithm with the approximate environment 
and the eeFU with the exact environment.
Fig.~\ref{fig:FU} shows that both algorithms give very similar magnetization plots.
The estimates of $T_c$ and $1/\tilde\beta\delta$ listed in Tab. \ref{tab:FU} are also the same within their 
error bars which are also similar.
We conclude that quality of the results is the same for both algorithms.

\subsection{Thermal states with simple update (SU) algorithm}
\label{sec:SU}

\begin{figure}[tb!]
\includegraphics[width=0.99\columnwidth,clip=true]{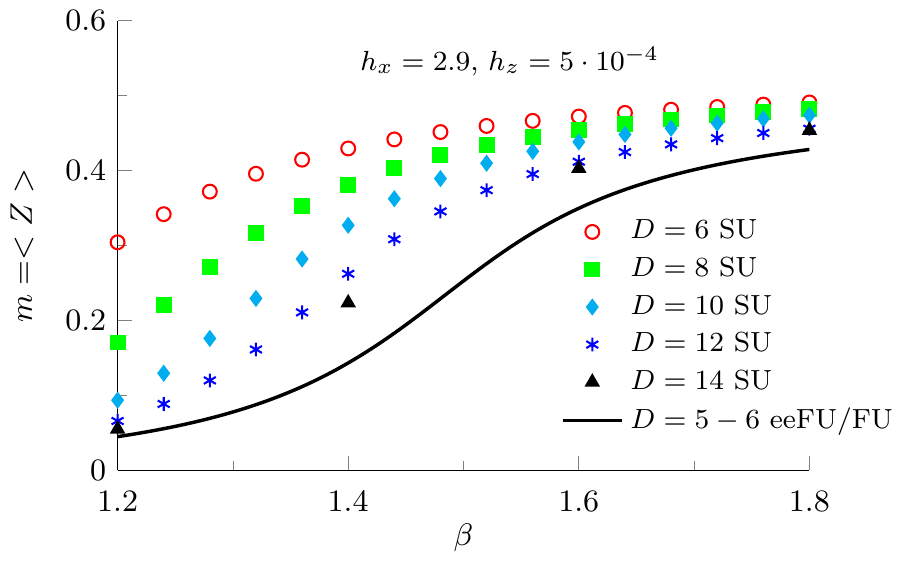}
\caption{
A comparison of thermal states obtained by the full update (FU) algorithm and the simple update (SU) evolution of 
a thermal state purification for $h_x=2.9$ and $h_z = 5\cdot 10^{-4}$. 
With increasing $D$ the SU magnetization moves slowly towards the converged FU magnetization ($D=5,6$)
but even for the largest $D=14$ it is still far from it.
}
\label{fig:SU}
\end{figure}

One can consider simplifying and accelerating the algorithm even further by replacing the full update (FU) 
of the PEPS tensors with the simple update (SU) \cite{Xiang_SU_08}. The SU truncates the enlarged bond dimension 
$kD$ by means of a singular value decomposition of the pair of PEPS tensors. Therefore, it ignores long range
correlations in the environment of the truncated bond.
The SU allows for larger $D$ 
because the bottleneck CTMRG procedure is needed only to compute observables in the final state. 
Recently thermal state simulation by the SU --
using time evolution of the density operator --
was proposed in Ref. \onlinecite{Orus_SUfiniteT_18}. 

\begin{table}[tb!]
\begin{tabular}{|l|c|c|c|}
\hline
 method  & $D$ &  $T_c$ &   $1/\tilde\beta\delta$ \\ 
\hline
SU &  $6$ & $0.867(2)$ & $1.04(7)$  \\
SU &  $8$ & $0.788(2)$ & $1.06(5)$  \\
SU & $10$ &  $0.741(3)$  & $1.00(5)$ \\
SU & $12$ &  $0.704(11)$  & $0.85(11)$ \\
FU  & $5$ & $0.6061(19)$ & $0.564(7)$ \\
QMC\cite{Hesselmann_TIsingQMC_16} & - & $0.6085(8)$ & - \\
exact  & - & - & $8/15 \approx 0.533$\\
\hline
\end{tabular}
\caption{Comparison of  simple update (SU) and FU estimates of $T_c$ and $1/\tilde\beta\delta$ obtained by the peak scaling (\ref{scalpeak}). Here $h_x=2.9$ and we use  $0.0005 \le h_z \le 0.01$ to perform the scaling. The $D=6-12$ SU results have much worse quality than $D=5$ FU results. Given that already for $D=12$ it takes longer to obtain the poor quality SU results than the good quality FU results with $D=5$, we conclude that the FU far outperforms the SU scheme.}
\label{tab:SU}
\end{table}

Here we compare the SU with the FU scheme.
Our SU algorithm is a straightforward generalization of the ground state algorithm\cite{Corboz_fiPEPS_10} 
to a purification of a thermal state.
In Fig.~\ref{fig:SU} we compare thermal states generated by both algorithms for $h_x=2.9$ and $h_z = 5\cdot 10^{-4}$.
With increasing $D$ the SU magnetization moves slowly towards the converged $D=5,6$ FU magnetization 
but even for $D=14$ it is still far from it. 
In Table \ref{tab:SU} we compare estimates of $T_c$ and $1/\tilde\beta\delta$ obtained for $h_x=2.9$ by the peak scaling (\ref{scalpeak}) with $0.0005 \le h_z \le 0.01$. Even for the largest $D=12$ the SU estimates are much worse than the FU estimates for $D=5$.
Given that already for $D=12$ SU requires more time than FU for $D=5$,
we conclude that the FU algorithm by far outperforms the SU algorithm, 
at least in the present example.

\section{Conclusion}
\label{sec:conclusion}

We tested efficient algorithms to simulate real, Lindbladian and imaginary time evolution with infinite PEPS. 
The key to the efficiency is local optimization of iPEPS tensors. 
In the case of imaginary time evolution of a thermal purification the accuracy can be improved by disentanglers applied 
to ancillas that reduce the necessary bond dimension. 
The efficiency can be enhanced further by reusing the tensor environment from the previous gate.
This simplification reduces the algorithm to the full update scheme.
We presented numerical evidence that this simplification does not affect the accuracy.
However,
further simplification to the simple update scheme is a step too far, at least for determining critical data in presence of strong quantum fluctuations. In such case the accuracy to determine the critical temperature deteriorates dramatically, when using the simple update.

A proof of principle demonstration was provided for unitary real time evolution after a sudden quench of the Hamiltonian,
Lindbladian evolution with a Markovian master equation, and imaginary time evolution generating thermal
states. In the last case, we used temperature dependence of the order parameter for different strengths of
the small symmetry-breaking bias field to estimate  critical temperature by extrapolation to the limit of vanishing bias field. 
We obtained a good accuracy of the critical temperature in the 2D quantum Ising model:
$0.1\%$ for the transverse field $h_x=2.5$ and $0.5 \%$ for the more challenging $h_x=2.9$ that is close to the quantum critical 
point at $h_x=3.0444$. 


\acknowledgments
We thank Stefan Wessel for numerical values of data published in Ref.~\onlinecite{Hesselmann_TIsingQMC_16} and Marek 
Rams for inspiring discussions. 
This research was funded by National Science Centre (NCN), Poland 
under project 2016/23/B/ST3/00830 (PC), NCN together with European Union through the QuantERA program 2017/25/Z/ST2/03028 (JD), and  the European Research Council (ERC) under the EU Horizon 2020 research and innovation program (Grant No. 677061).

\appendix

\begin{figure}[tb!]
\includegraphics[width=0.99\columnwidth,clip=true]{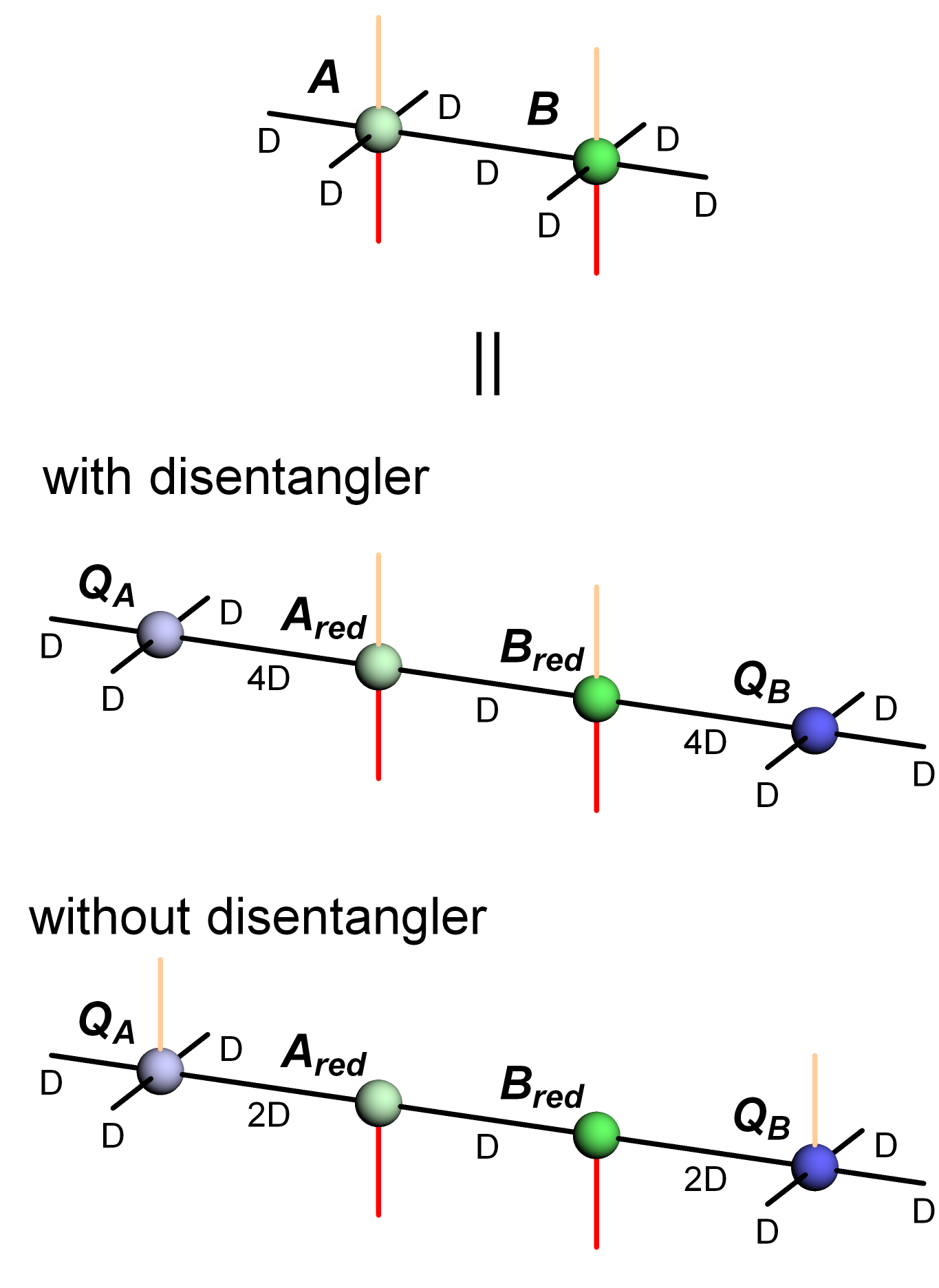}
\caption{
Tensors $A$ and $B$ are decomposed into isometries $Q_A$ and $Q_B$ and smaller
reduced tensors $A_{\rm red}$ and $B_{\rm red}$. A similar decomposition,
with the same $Q_A$ and $Q_B$, 
applies to the pairs $A',B'$ and $A'',B''$.
We show two versions of  reduced tensors.  
The larger ones are intended for simulations with disentanglers while the smaller ones, 
which leave ancilla indices with the isometries, 
for simulations without the disentanglers.  
}
\label{fig:reduced}
\end{figure}

\section{Reduced tensors} 
\label{app:reduced}

For the sake of clarity, in the main body of the paper the algorithms were presented with full tensors $A$ and $B$.
In our actual numerical calculations, however, we optimize reduced tensors, see Fig.~\ref{fig:reduced}.
The full tensor $A$ ($B$) is a contraction of an isometry $Q_A$ ($Q_B$) with a reduced tensor
$A_{\rm red}$ ($B_{\rm red}$). It is obtained with the help of QR decomposition
\be 
A= Q_A A_{\rm red}, 
\ee 
were $Q_A$ is an isometry and $A_{\rm red}$ is an upper triangular matrix.  
For spin $1/2$ and with disentangler, 
$A_{\rm red}$ has $16D^2$ elements instead of $4D^4$ elements of the full tensor $A$. 
In the case without the disentanglers 
one can also use the smaller reduced tensors with just $4D^2$ elements.

In the local optimization procedure isometries $Q_A$ and $Q_B$ are held constant:
\bea 
A'  &=& Q_A A'_{\rm red},~~ A'' = Q_A A''_{\rm red}, \\
B'  &=& Q_B B'_{\rm red},~~ B'' = Q_B B''_{\rm red}. 
\eea 
Rather than full tensors $A''$ and $B''$, only $A''_{\rm red}$ and $B''_{\rm red}$ are subject to optimization in the loop (\ref{loop}). 
We note that the reduced tensors are commonly used in ground state iPEPS simulations, see 
e. g. Ref. \onlinecite{Corboz_fiPEPS_10}. 

For $D>2$ using the reduced tensors we decrease cost of the tensor optimization (\ref{loop}). The larger reduced tensors are necessary for simulations with the disentanglers. Our numerical tests suggest that the smaller ones are better for simulations without disentanglers 
as they provide the same accuracy as the larger ones while reducing the cost of the optimization loop.
However, in section \ref{sec:dis} we use the larger reduced tensors both with and without disentanglers to make the comparison
more reliable.

\section{Numerical details} 
\label{app:num}

As a criterion of CTMRG convergence we use change of the reduced tensors 2-site environment. We demand relative change of its 2-norm per iteration smaller than $10^{-10}$. The time cost of obtaining  full update  $T_c$ and $1/\tilde\beta\delta$ estimates for $h_x=2.9$ and $D=5$ shown in Tab. \ref{tab:FU} was 5-6 days using a 14 core, 2.20 GHz, Intel Xeon Gold 5120 processor.

\bibliography{exact.bib} 
\end{document}